\documentclass{article}
\usepackage{spconf,amsmath,graphicx,bm,booktabs,standalone,tikz,hyperref}
\usetikzlibrary{calc, trees, positioning, arrows, shapes, fit, patterns, backgrounds}
\graphicspath{{./imgs/}}

\title{Multiscale Augmented Normalizing Flows for Image Compression}
\name{Marc Windsheimer, Fabian Brand, Andr\'e Kaup\thanks{This work was partly funded by the Deutsche Forschungsgemeinschaft (DFG, German Research Foundation) - Project ID 426084215.}}
\address{Multimedia Communications and Signal Processing\\ Friedrich-Alexander-Universit\"at Erlangen-N\"urnberg\\ Cauerstr. 7, 91058 Erlangen, Germany}
\newcommand\copyrighttext{%
	\footnotesize \textcopyright 2023 IEEE. Personal use of this material is permitted.
	Permission from IEEE must be obtained for all other uses, in any current or future 
	media, including reprinting/republishing this material for advertising or promotional 
	purposes, creating new collective works, for resale or redistribution to servers or 
	lists, or reuse of any copyrighted component of this work in other works. DOI: \href{https://doi.org/10.1109/ICASSP48485.2024.10446147}{10.1109/ICASSP48485.2024.10446147}

}
\newcommand\copyrightnoticeOwn{%
	\begin{tikzpicture}[remember picture,overlay]
		\node[anchor=north,yshift=-10pt] at (current page.north) {\fbox{\parbox{\dimexpr\textwidth-\fboxsep-\fboxrule\relax}{\copyrighttext}}};
	\end{tikzpicture}%
	\vspace{-8mm}
}

\begin{document}
	\ninept
	\maketitle
	\copyrightnoticeOwn
	\begin{abstract}
		Most learning-based image compression methods lack efficiency for high image quality due to their non-invertible design.
		The decoding function of the frequently applied compressive autoencoder architecture is only an approximated inverse of the encoding transform.
		This issue can be resolved by using invertible latent variable models, which allow a perfect reconstruction if no quantization is performed.
		Furthermore, many traditional image and video coders apply dynamic block partitioning to vary the compression of certain image regions depending on their content.
		Inspired by this approach, hierarchical latent spaces have been applied to learning-based compression networks.
		In this paper, we present a novel concept, which adapts the hierarchical latent space for augmented normalizing flows, an invertible latent variable model.
		Our best performing model achieves significant rate savings of more than 7\% over comparable single-scale models.
	\end{abstract}
	\begin{keywords}
		Learning-based image compression, augmented normalizing flows, hierarchical image compression, rate distortion optimization, end-to-end image compression
	\end{keywords}
	\section{Introduction}
	\label{sec:intro}
	In recent years, learning-based image compression has gained much attention, as it surpassed state-of-the-art image and video coding standards, e.g. BPG or VVC \cite{VVC}.
	These standards rely on hand-crafted features to achieve a suitable trade-off between compression efficiency and complexity.
	In contrast, learning-based compression methods are optimized in an end-to-end fashion to jointly learn the parameters of the encoding and decoding function.
	Compressive autoencoders (CAE) \cite{Balle2017} have become a common approach for end-to-end learned image coding.
	The non-linear encoding function computes a latent representation depending on the original image, which is then quantized and encoded, typically by arithmetic coding.
	The quantized latent space is fed into the decoding function to generate a reconstructed version of the original image.
	Current research mainly focuses on optimizing the encoding/decoding transformations and the entropy models for the accurate prediction of the likelihood of the latent space symbols \cite{He2021,Fu2023,Liu2023}.
	
	A downside of most learning-based image compression methods, including CAE-based architectures, is their performance for high bit rates due to the lack of invertibility.
	Even if quantization is omitted, the reconstructed image is not completely identical to the original image, since the decoding function is not a perfect inverse of the encoding transform.
	The use of invertible latent variable models, like Augmented Normalizing Flows (ANF) \cite{Huang2020}, can improve the performance of learning-based image compression, especially for high quality image coding.
	In terms of lossy compression, this invertible design can reduce the saturation effects for high bitrates, which typically occur with CAE-based architectures.
	
	Furthermore, learning-based image compression models are usually fully determined after the training and cannot adapt to the image content.
	In contrast, state-of-the-art traditional image and video coding standards apply adaptive block partitioning to modify the coding structure depending on the image content.
	This drawback has been addressed by RDONet \cite{Brand2021,Brand2022a,Brand2022b} with the application of a hierarchical latent space.
	This hierarchical design allows the network to transmit the latent representation for certain image areas at variable scales.
	While highly structured regions are typically transmitted using small blocks to preserve high quality, larger blocks are used for less important areas to reduce the bit rate necessary for coding the corresponding image region.
	The decision in which hierarchy level a certain image area will be transmitted can be set during inference and thus, the network can adjust its behaviour even after the learned parameters have been fixed.
	Since the larger block sizes are especially beneficial in the low-rate regime, the compression performance can be significantly improved.
	
	In this work, the ANF-based architecture ANFIC \cite{Ho2021b} is extended with a multiscale latent space.
	The invertible design allows image compression at very high quality, while the adaptive latent space can reduce the necessary bit rate.
	
	\section{Related Work}
	\label{sec:related}
	\begin{figure*}
		\centering
			\begin{tikzpicture}[
		dataBlob/.style={rectangle, anchor = center},
		Layer/.style={rectangle, draw, anchor = center, fill=green!20, minimum width=2.5cm},
		Attention/.style={rectangle, draw, anchor = center, fill=yellow!20},
		Activation/.style={rectangle, draw, anchor = center, fill=blue!20},
		Misc/.style={rectangle, draw, anchor = center, fill=black!20},
		Code/.style={, draw, anchor = center, fill=black!20},
		bg/.style={rectangle, anchor = center, fill=\colorBG, anchor = north west,},
		myArrow/.style={-stealth},
		myArrowG/.style={*-stealth},
		myArrowL/.style={open diamond-stealth},
		myArrowRE/.style={diamond-stealth},
		myArrowR/.style={stealth-},
		label/.style={rectangle, anchor = center},
		operator/.style={circle ,draw, anchor = center, minimum size = 0.2cm, inner sep = 0pt},
		Network/.style={rectangle, anchor = north west, fill = blue!10, draw = blue!60},
		]
		
		\node (x) [dataBlob] at (0,0) {\Large$\bm{x}$};
		
		\node (dummy1) [label, right of=x, node distance=2cm] {};
		\node (enc01) [Layer, below of=dummy1, node distance=0.75cm] {Conv $L$/5/2$\downarrow$};
		\node (enc02) [Layer, below of=enc01, node distance=0.9cm] {Conv $L$/5/2$\downarrow$};
		\node (enc03) [Layer, below of=enc02, node distance=0.9cm] {Conv $L$/5/2$\downarrow$};
		
		\node (plus0) [operator, right of=dummy1, node distance=3cm] {+};
		\node (dec01) [Layer, below of=plus0, node distance=0.75cm] {TConv 3/5/2$\uparrow$};
		\node (dec02) [Layer, below of=dec01, node distance=0.9cm] {TConv $L$/5/2$\uparrow$};
		\node (dec03) [Layer, below of=dec02, node distance=0.9cm] {TConv $L$/5/2$\uparrow$};
		\node (minus0) [label, below right=-0.1cm and -0.1cm of plus0] {$\boldsymbol{-}$};
		
		\node (dummy3) [label] at ($(enc03)!0.5!(dec03)$) {};
		\node (LSUnit01) [Network, below of=dummy3, node distance=0.9cm, minimum width=5.5cm] {LS-Unit (Type A)};
		\node (LSUnit02) [Network, below of=LSUnit01, node distance=0.9cm, minimum width=5.5cm] {LS-Unit 2 (Type A)};
		
		\node (x1) [dataBlob, right of=plus0, node distance=2cm] {\Large$\bm{x}_1$};
		
		\node (dummy4) [label, right of=x1, node distance=2cm] {};
		\node (enc11) [Layer, below of=dummy4, node distance=0.75cm] {Conv $L$/5/2$\downarrow$};
		\node (enc12) [Layer, below of=enc11, node distance=0.9cm] {Conv $L$/5/2$\downarrow$};
		\node (enc13) [Layer, below of=enc12, node distance=0.9cm] {Conv $L$/5/2$\downarrow$};
		
		\node (plus1) [operator, right of=dummy4, node distance=3cm] {+};
		\node (dec11) [Layer, below of=plus1, node distance=0.75cm] {TConv 3/5/2$\uparrow$};
		\node (dec12) [Layer, below of=dec11, node distance=0.9cm] {TConv $L$/5/2$\uparrow$};
		\node (dec13) [Layer, below of=dec12, node distance=0.9cm] {TConv $L$/5/2$\uparrow$};
		\node (minus1) [label, below right=-0.1cm and -0.1cm of plus1] {$\boldsymbol{-}$};
		
		\node (dummy6) [label] at ($(enc13)!0.5!(dec13)$) {};
		\node (LSUnit11) [Network, below of=dummy6, node distance=0.9cm, minimum width=5.5cm] {LS-Unit 1 (Type B)};
		\node (LSUnit12) [Network, below of=LSUnit11, node distance=0.9cm, minimum width=5.5cm] {LS-Unit 2 (Type B)};
		
		\node (x2) [dataBlob, right of=plus1, node distance=2cm] {\Large$\bm{x}_2$};
		
		\node (zero0) [dataBlob, left of=LSUnit01, node distance=3.5cm] {\Large$\bm{0}$};
		\node (zero1) [dataBlob, left of=LSUnit02, node distance=3.5cm] {\Large$\bm{0}$};
		\node (z11) [dataBlob] at ($(LSUnit01)!0.5!(LSUnit11)$) {\Large$\bm{z}_{11}$};
		\node (z12) [dataBlob] at ($(LSUnit02)!0.5!(LSUnit12)$) {\Large$\bm{z}_{12}$};
		\node (z21) [dataBlob, right of=LSUnit11, node distance=3.5cm] {\Large$\hat{\bm{z}}_{21}$};
		\node (z22) [dataBlob, right of=LSUnit12, node distance=3.5cm] {\Large$\hat{\bm{z}}_{22}$};
		
		\begin{pgfonlayer}{background}
			\node (shade) [pattern=north west lines, pattern color=gray!70, fit=(zero0) (z22), inner sep=5pt] {};
		\end{pgfonlayer}
		
		\draw[myArrow] (x) -- (plus0);
		\draw[myArrow] (plus0) -- (x1);
		\draw[myArrow] (x1) -- (plus1);
		\draw[myArrow] (plus1) -- (x2);
		
		\draw[myArrow] (dummy1.center) -- (enc01);
		\draw[myArrowG] (enc01) -- (enc02);
		\draw[myArrowG] (enc02) -- (enc03);
		\draw[myArrowG] (enc03) -- ([xshift=-1.5cm] LSUnit01.north);
		\draw[myArrow] ([xshift=-1.5cm] LSUnit01.south) -- ([xshift=-1.5cm] LSUnit02.north);
		
		\draw[myArrow] ([xshift=1.5cm] LSUnit02.north) -- ([xshift=1.5cm] LSUnit01.south);
		\draw[myArrow] ([xshift=1.5cm] LSUnit01.north) -- (dec03);
		\draw[myArrowG] (dec03) -- (dec02);
		\draw[myArrowG] (dec02) -- (dec01);
		\draw[myArrow] (dec01) -- (plus0);
		
		\draw[myArrow] (dummy4.center) -- (enc11);
		\draw[myArrowG] (enc11) -- (enc12);
		\draw[myArrowG] (enc12) -- (enc13);
		\draw[myArrowG] (enc13) -- ([xshift=-1.5cm] LSUnit11.north);
		\draw[myArrow] ([xshift=-1.5cm] LSUnit11.south) -- ([xshift=-1.5cm] LSUnit12.north);
		
		\draw[myArrow] ([xshift=1.5cm] LSUnit12.north) -- ([xshift=1.5cm] LSUnit11.south);
		\draw[myArrow] ([xshift=1.5cm] LSUnit11.north) -- (dec13);
		\draw[myArrowG] (dec13) -- (dec12);
		\draw[myArrowG] (dec12) -- (dec11);
		\draw[myArrow] (dec11) -- (plus1);
		
		\draw[myArrow] (zero0) -- (LSUnit01);
		\draw[myArrow] (LSUnit01) -- (z11);
		\draw[myArrow] (z11) -- (LSUnit11);
		\draw[myArrow] (LSUnit11) -- (z21);
		\draw[myArrow] (zero1) -- (LSUnit02);
		\draw[myArrow] (LSUnit02) -- (z12);
		\draw[myArrow] (z12) -- (LSUnit12);
		\draw[myArrow] (LSUnit12) -- (z22);
		
	\end{tikzpicture}
		\caption{Architecture overview of \mbox{M-ANFIC}. \textit{Conv} $c/k/s\downarrow$ denotes a convolution layer with $c$ output channels, a kernel size of $k\times k$ and a downsampling factor of $s$. \textit{TConv} $c/k/s\uparrow$ denotes an identical parametrized transposed convolution. Filled circles after a convolution represent GDNs/IGDNs \cite{Balle2016}. Shaded area marks the novel components.}
		\label{fig:m_anfic_architecture}
	\end{figure*}
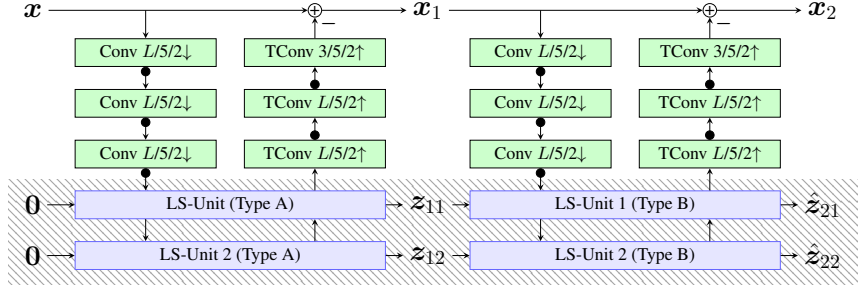
	The idea of a compressive autoencoder for image coding has been proposed in \cite{Balle2017}.
	Up to now, multiple enhancements have been developed, extending the initial architecture.
	A major improvement is the scale hyperprior \cite{Balle2018}, which derives the parameters of a density model for the latent space symbols from additional side-information.
	By assigning shorter code words to more likely symbols, the benefit of saved bits outweighs the necessary rate for transmitting the additional side-information.
	This idea has been further improved by predicting also the mean in combination with an autoregressive context model \cite{Minnen2018} or using Gaussian mixture models (GMM) as density model \cite{Cheng2020}.
	
	The work within this paper is based on two distinct architectures for image compression. 
	Namely, ANFIC \cite{Ho2021b} and \mbox{RDONet \cite{Brand2022b}}.
	The ANFIC architecture applies augmented normalizing flows \cite{Huang2020} with two autoencoding layers, where the latent space is hierarchically augmented by a combination of a hyperprior and an autoregressive context model.
	Similar to \cite{Cheng2020}, ANFIC uses a GMM with $K=3$ components to model the likelihoods of the latent space symbols.
	Since quantization is applied and the residual after encoding is replaced by zeros for the decoding step, the ANFIC architecture is still a lossy compression method.
	Due to the invertible network design, the introduced errors cannot be concealed during reconstruction.
	Therefore, the output after decoding is fed into a quality enhancement network \cite{Ma2022}, which reduces the visible effects of the performed quantization.
	
	RDONet \cite{Brand2021} is a traditional compressive autoencoder based \mbox{on \cite{Minnen2018}}.
	The main novelty of RDONet is the hierarchical latent space, which allows to transmit the latent representation in either of the multiple latent space units (LSUnit).
	The decision in which LSUnit a certain region of the image will be transmitted can be set during inference.
	Thus, the model preserves a certain degree of freedom to optimize the partitioning of the block size using Rate Distortion Optimization (RDO).
	The overall image is then reconstructed based on the transmitted latent in all LSUnits.
	Initially, RDONet used solely random masks during training to decide where a certain image region is transmitted.
	In later publications \cite{Brand2022a,Brand2022b} however, the training uses masks based on a variance criterion after a certain amount of training epochs.
	Furthermore, these variance-based masks have proven to be a good estimation for a suitable block partitioning during inference and can be an alternative to the time-consuming RDO \cite{Brand2022a}.
	
	\section{Proposed Method}
	\label{sec:multiscaleanfic}
	In this section, we propose two novel architectures, which extend ANFIC by a multiscale approach, which is inspired by the work on RDONet.
	Identical to the ANFIC design, our \mbox{M-ANFIC} and \mbox{MS-ANFIC} models apply augmented normalizing flows with two autoencoding layers.
	
	\subsection{M-ANFIC}
	\label{subsec:m-anfic}
	\begin{figure}
		\centering
			\begin{tikzpicture}[
		dataBlob/.style={rectangle, anchor = center},
		Layer/.style={rectangle, draw, anchor = center, fill=green!20, minimum width=2.5cm},
		Attention/.style={rectangle, draw, anchor = center, fill=yellow!20},
		Activation/.style={rectangle, draw, anchor = center, fill=blue!20},
		Misc/.style={rectangle, draw, anchor = center, fill=black!20},
		Code/.style={, draw, anchor = center, fill=black!20},
		bg/.style={rectangle, anchor = center, anchor = north west,},
		myArrow/.style={-stealth},
		myArrowG/.style={*-stealth},
		myArrowL/.style={open diamond-stealth},
		myArrowRE/.style={diamond-stealth},
		myArrowR/.style={stealth-},
		label/.style={rectangle, anchor = center},
		operator/.style={circle ,draw, anchor = center, minimum size = 0.2cm, inner sep = 0pt},
		Network/.style={rectangle, anchor = north west, fill = blue!10, draw = blue!60},
		]
		
		\node (inputE) [dataBlob] at (0, 0) {\Large$\bm{u}_n$};
		\node (enc) [Layer, below of=inputE, node distance=0.9cm] {Conv $L$/5/2$\downarrow$};
		\node (outputD) [dataBlob, right of=inputE, node distance=3cm] {\Large$\bm{v}_n$};
		\node (dec) [Layer, below of=outputD, node distance=0.9cm] {TConv $L$/5/2$\uparrow$};
		\node (dummy1) [label] at ($(enc)!0.5!(dec)$) {};
		\node (plus) [operator, below of=dummy1, node distance=0.75cm] {\Large$+$};
		\node (inputZ) [dataBlob, left of=plus, node distance=3.5cm] {\Large$\bm{z}_{0n}$};
		\node (outputZ) [dataBlob, right of=plus, node distance=3.6cm] {\Large$\bm{z}_{1n}$};
		\node (mask) [Misc, below of=plus, node distance=0.75cm] {Masking};
		\node (inner) [Layer, below of=mask, node distance=0.75cm] {Conv $N$/3/1$\downarrow$};
		\node (dummy2) [label, below of=inner, node distance=0.5cm] {};
		\node (outputE) [dataBlob, below of=enc, node distance=3.5cm] {\Large$\bm{u}_{n+1}$};
		\node (inputD) [dataBlob, below of=dec, node distance=3.5cm] {\Large$\bm{v}_{n+1}$};
		
		\begin{pgfonlayer}{background}
			\node (bg) [bg, draw=blue!60, fill=blue!10, fit=(enc) (dec) (dummy2), inner sep=5pt, yshift=1pt] {};
		\end{pgfonlayer}

		\draw[myArrow] (inputE) -- (enc);
		\draw[myArrowG] (enc) -- (outputE);
		\draw[myArrow] ([yshift=0.75cm] outputE.center) -- ([xshift=-0.25cm] dummy2.center) -- ([xshift=-0.25cm] inner.south);
		\draw[myArrow] ([yshift=0.75cm] inputD.center) -- ([xshift=0.25cm] dummy2.center) -- ([xshift=0.25cm] inner.south);
		\draw[myArrow] (inputD) -- (dec);
		\draw[myArrowG] (dec) -- (outputD);
		\draw[myArrow] (inner) -- (mask);
		\draw[myArrow] (mask) -- (plus.south);
		\draw[thick, densely dashed, -stealth, red] (plus.east) -| ([xshift=-0.5cm] dec.south);
		\draw[thick, densely dashed, -stealth, red] (inputZ) -- (plus.west);
		\draw[thick, densely dashed, -stealth, red] (plus.east) -- (outputZ);
		
	\end{tikzpicture}
		\caption{Structure of LSUnit (Type A). Notation analogous to Fig. \ref{fig:m_anfic_architecture}. Two signals entering the same convolution corresponds to a concatenation along the channel dimension. Latent space is masked according to the external mask.}
		\label{fig:lsunit_A}
	\end{figure}
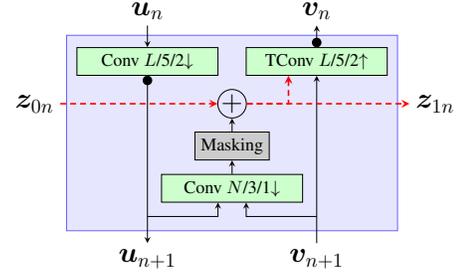
	
	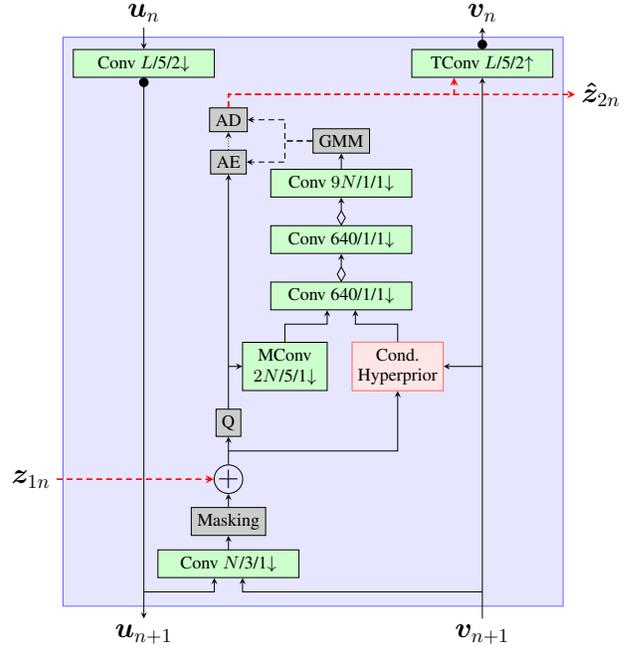
\begin{figure}
		\centering
			\begin{tikzpicture}[
		dataBlob/.style={rectangle, anchor = center},
		Layer/.style={rectangle, draw, anchor = center, fill=green!20, minimum width=2.5cm},
		Attention/.style={rectangle, draw, anchor = center, fill=yellow!20},
		Activation/.style={rectangle, draw, anchor = center, fill=blue!20},
		Misc/.style={rectangle, draw, anchor = center, fill=black!20},
		Code/.style={, draw, anchor = center, fill=black!20},
		bg/.style={rectangle, anchor = center, anchor = north west,},
		myArrow/.style={-stealth},
		myArrowG/.style={*-stealth},
		myArrowL/.style={open diamond-stealth},
		myArrowRE/.style={diamond-stealth},
		myArrowR/.style={stealth-},
		label/.style={rectangle, anchor = center},
		operator/.style={circle ,draw, anchor = center, minimum size = 0.2cm, inner sep = 0pt},
		Network/.style={rectangle, anchor = north west, fill = blue!10, draw = blue!60},
		]
		
		\node (inputE) [dataBlob] at (0, 0) {\Large$\bm{u}_n$};
		\node (enc) [Layer, below of=inputE, node distance=0.9cm] {Conv $L$/5/2$\downarrow$};
		\node (outputD) [dataBlob, right of=inputE, node distance=6cm] {\Large$\bm{v}_n$};
		\node (dec) [Layer, below of=outputD, node distance=0.9cm] {TConv $L$/5/2$\uparrow$};
		\node (AD) [Misc, below of=enc, xshift=1.5cm] {AD};
		\node (AE) [Misc, below of=AD, node distance=0.75cm] {AE};
		\node (dummy1) [label, xshift=1cm] at ($(AD)!0.5!(AE)$) {};
		\node (GMM) [Misc, right of=dummy1] {GMM};
		\node (agg3) [Layer, below of=GMM, node distance=0.75cm] {Conv $9N$/1/1$\downarrow$};
		\node (agg2) [Layer, below of=agg3] {Conv 640/1/1$\downarrow$};
		\node (agg1) [Layer, below of=agg2] {Conv 640/1/1$\downarrow$};
		\node (MConv) [Layer, below of=agg1, node distance=1.25cm, xshift=-1cm, align=center, minimum width=1.5cm] {MConv \\ $2N$/5/1$\downarrow$};
		\node (hyp) [Network, below of=agg1, node distance=1.25cm, xshift=1cm, align=center, minimum width=1.5cm, fill=red!10, draw=red!60] {Cond. \\ Hyperprior};
		\node (Q) [Misc, below of=MConv, xshift=-1cm] {Q};
		\node (plus) [operator, below of=Q] {\Large$+$};
		\node (inputZ) [dataBlob, left of=plus, node distance=3.5cm] {\Large$\bm{z}_{1n}$};
		\node (outputZ) [dataBlob, right of=AD, node distance=6.6cm, yshift=0.45cm] {\Large$\bm{\hat{z}}_{2n}$};
		\node (mask) [Misc, below of=plus, node distance=0.75cm] {Masking};
		\node (inner) [Layer, below of=mask, node distance=0.75cm] {Conv $N$/3/1$\downarrow$};
		\node (dummy2) [label, below of=inner, node distance=0.5cm] {};
		\node (outputE) [dataBlob, below of=enc, node distance=10.125cm] {\Large$\bm{u}_{n+1}$};
		\node (inputD) [dataBlob, below of=dec, node distance=10.125cm] {\Large$\bm{v}_{n+1}$};
		
		\begin{pgfonlayer}{background}
			\node (bg) [bg, draw=blue!60, fill=blue!10, fit=(enc) (dec) (dummy2), inner sep=5pt, yshift=1pt] {};
		\end{pgfonlayer}
	
		\draw[myArrow] (inputE) -- (enc);
		\draw[myArrowG] (enc) -- (outputE);
		\draw[myArrow] ([yshift=0.75cm] outputE.center) -- ([xshift=-0.25cm] dummy2.center) -- ([xshift=-0.25cm] inner.south);
		\draw[myArrow] ([yshift=0.75cm] inputD.center) -- ([xshift=0.25cm] dummy2.center) -- ([xshift=0.25cm] inner.south);
		\draw[myArrow] (inputD) -- (dec);
		\draw[myArrowG] (dec) -- (outputD);
		\draw[myArrow] (inner) -- (mask);
		\draw[myArrow] (mask) -- (plus.south);
		\draw[myArrow] ([yshift=1cm] Q.center) -- (MConv);
		\draw[myArrow] ([yshift=0.5cm] plus.center) -| (hyp);
		\draw[myArrow] ([xshift=1.5cm] hyp.center) -- (hyp);
		\draw[myArrow] (MConv) |- ([xshift=-0.25cm, yshift=-0.5cm] agg1.center) -- ([xshift=-0.25cm] agg1.south);
		\draw[myArrow] (hyp) |- ([xshift=0.25cm, yshift=-0.5cm] agg1.center) -- ([xshift=0.25cm] agg1.south);
		\draw[myArrowL] (agg1) -- (agg2);
		\draw[myArrowL] (agg2) -- (agg3);
		\draw[myArrow] (agg3) -- (GMM);
		\draw[myArrow] (Q) -- (AE);
		\draw[myArrow, densely dashed] (GMM) -- ([xshift=-1cm] GMM.center) |- (AE);
		\draw[myArrow, densely dashed] (GMM) -- ([xshift=-1cm] GMM.center) |- (AD);
		\draw[densely dotted, -stealth] (AE) -- (AD);
		\draw[thick, densely dashed, -stealth, red] (AD) |- (outputZ);
		\draw[thick, densely dashed, -stealth, red] (AD) |- ([xshift=-0.5cm, yshift=-0.55cm] dec.center) -- ([xshift=-0.5cm] dec.south);
		\draw[thick, densely dashed, -stealth, red] (inputZ) -- (plus.west);
		\draw[myArrow] (plus.north) -- (Q);
		
	\end{tikzpicture}
		\caption{Structure of LSUnit (Type B). Notation analogous to Figs. \ref{fig:m_anfic_architecture} and \ref{fig:lsunit_A}. Blank rhombuses denote leaky ReLUs. The architecture of the conditional hyperprior is identical to the hyperprior of \mbox{ANFIC \cite{Ho2021b}} but with $\bm{v}_{n+1}$ concatenated to the input of the first encoding and last decoding layer.}
		\label{fig:lsunit_B}
	\end{figure}
	
	\begin{figure*}
		\centering
			\begin{tikzpicture}[
		dataBlob/.style={rectangle, anchor = center},
		Layer/.style={rectangle, draw, anchor = center, fill=green!20, minimum width=2.5cm},
		Attention/.style={rectangle, draw, anchor = center, fill=yellow!20},
		Activation/.style={rectangle, draw, anchor = center, fill=blue!20},
		Misc/.style={rectangle, draw, anchor = center, fill=black!20},
		Code/.style={, draw, anchor = center, fill=black!20},
		bg/.style={rectangle, anchor = center, fill=\colorBG, anchor = north west,},
		myArrow/.style={-stealth},
		myArrowG/.style={*-stealth},
		myArrowL/.style={open diamond-stealth},
		myArrowRE/.style={diamond-stealth},
		myArrowR/.style={stealth-},
		label/.style={rectangle, anchor = center},
		operator/.style={circle ,draw, anchor = center, minimum size = 0.2cm, inner sep = 0pt},
		Network/.style={rectangle, anchor = north west, fill = blue!10, draw = blue!60},
		]

		\node (x) [dataBlob] at (0,0) {\Large$\bm{x}$};
		\node (dummy1) [label, right of=x, node distance=2cm] {};
		\node (enc01) [Layer, below of=dummy1, node distance=0.75cm] {Conv $L$/5/2$\downarrow$};
		\node (enc02) [Layer, below of=enc01, node distance=0.9cm] {Conv $L$/5/2$\downarrow$};
		\node (enc03) [Layer, below of=enc02, node distance=0.9cm] {Conv $L$/5/2$\downarrow$};
		\node (enc04) [Layer, below of=enc03, node distance=0.9cm] {Conv $N$/5/2$\downarrow$};
		\node (plus1) [operator, below of=enc04, node distance=0.75cm] {+};
		\node (zeros) [label, left of=plus1, node distance=2cm] {\Large$\bm{0}$};
		\node (z1) [label, right of=plus1, node distance=1.5cm] {\Large$\bm{z}_1$};
		
		\node (plus2) [operator, right of=dummy1, node distance=3cm] {+};
		\node (minus1) [label, below right=-0.1cm and -0.1cm of plus2] {$\boldsymbol{-}$};
		\node (dec01) [Layer, below of=plus2, node distance=0.75cm] {TConv 3/5/2$\uparrow$};
		\node (dec02) [Layer, below of=dec01, node distance=0.9cm] {TConv $L$/5/2$\uparrow$};
		\node (dec03) [Layer, below of=dec02, node distance=0.9cm] {TConv $L$/5/2$\uparrow$};
		\node (dec04) [Layer, below of=dec03, node distance=0.9cm] {TConv $L$/5/2$\uparrow$};
		\node (dummy4) [label, right of=z1, node distance=1.5cm] {};
		
		\node (lssplit) [Network, right of=dummy4, node distance=3.25cm, fill=orange!10, draw=orange!60, minimum width=3cm, minimum height=1.5cm, align=center, yshift=0.3cm] {Latent Split\\Network};
		
		\node (x1) [dataBlob, right of=plus2, node distance=3.75cm] {\Large$\bm{x}_1$};
		
		\node (dummy5) [label, right of=x1, node distance=3.75cm] {};
		\node (enc11) [Layer, below of=dummy5, node distance=0.75cm] {Conv $L$/5/2$\downarrow$};
		\node (enc12) [Layer, below of=enc11, node distance=0.9cm] {Conv $L$/5/2$\downarrow$};
		\node (enc13) [Layer, below of=enc12, node distance=0.9cm] {Conv $L$/5/2$\downarrow$};
		
		\node (plus3) [operator, right of=dummy5, node distance=3cm] {+};
		\node (minus2) [label, below right=-0.1cm and -0.1cm of plus3] {$\boldsymbol{-}$};
		\node (dec11) [Layer, below of=plus3, node distance=0.75cm] {TConv 3/5/2$\uparrow$};
		\node (dec12) [Layer, below of=dec11, node distance=0.9cm] {TConv $L$/5/2$\uparrow$};
		\node (dec13) [Layer, below of=dec12, node distance=0.9cm] {TConv $L$/5/2$\uparrow$};

		\node (dummy6) [label] at ($(enc13)!0.5!(dec13)$) {};
		\node (LSUnit11) [Network, below of=dummy6, node distance=0.9cm, minimum width=5.5cm] {LS-Unit 1 (Type B)};
		\node (LSUnit12) [Network, below of=LSUnit11, node distance=0.9cm, minimum width=5.5cm] {LS-Unit 2 (Type B)};
		
		\node (z11) [label, left of=LSUnit11, node distance=3.5cm] {\Large$\bm{z}_{11}$};
		\node (z12) [label, left of=LSUnit12, node distance=3.5cm] {\Large$\bm{z}_{12}$};
		
		\node (x2) [dataBlob, right of=plus3, node distance=2cm] {\Large$\bm{x}_2$};
		\node (z21) [label, right of=LSUnit11, node distance=3.5cm] {\Large$\hat{\bm{z}}_{21}$};
		\node (z22) [label, right of=LSUnit12, node distance=3.5cm] {\Large$\hat{\bm{z}}_{22}$};
		
		\begin{pgfonlayer}{background}
			\node (shade) [pattern=north west lines, pattern color=gray!70, fit=(lssplit) (z21) (z22), inner sep=5pt] {};
		\end{pgfonlayer}
		
		\draw[myArrow] (zeros) -- (plus1);
		\draw[myArrow] (plus1) -- (z1);
		\draw[myArrow] (x) -- (plus2);
		\draw[myArrow] (plus2) -- (x1);
		\draw[myArrow] (dummy1.center) -- (enc01);
		\draw[myArrowG] (enc01) -- (enc02);
		\draw[myArrowG] (enc02) -- (enc03);
		\draw[myArrowG] (enc03) -- (enc04);
		\draw[myArrow] (enc04) -- (plus1);
		\draw[myArrow] (z1) -- ([yshift=-0.3cm] lssplit.west);
		\draw[myArrow] (dummy4.center) -- (dec04);
		\draw[myArrowG] (dec04) -- (dec03);
		\draw[myArrowG] (dec03) -- (dec02);
		\draw[myArrowG] (dec02) -- (dec01);
		\draw[myArrow] (dec01) -- (plus2);
		
		\draw[myArrow] (x1) -- (plus3);
		\draw[myArrow] (plus3) -- (x2);
		\draw[myArrow] (dummy5.center) -- (enc11);
		\draw[myArrowG] (enc11) -- (enc12);
		\draw[myArrowG] (enc12) -- (enc13);
		\draw[myArrowG] (enc13) -- ([xshift=-1.5cm] LSUnit11.north);
		\draw[myArrow] ([xshift=-1.5cm] LSUnit11.south) -- ([xshift=-1.5cm] LSUnit12.north);
		\draw[myArrow] ([xshift=1.5cm] LSUnit12.north) -- ([xshift=1.5cm] LSUnit11.south);
		\draw[myArrow] ([xshift=1.5cm] LSUnit11.north) -- (dec13);
		\draw[myArrowG] (dec13) -- (dec12);
		\draw[myArrowG] (dec12) -- (dec11);
		\draw[myArrow] (dec11) -- (plus3);
		
		\draw[myArrow] ([yshift=0.45cm] lssplit.east) -- (z11);
		\draw[myArrow] ([yshift=-0.45cm] lssplit.east) -- (z12);
		\draw[myArrow] (z11) -- (LSUnit11);
		\draw[myArrow] (z12) -- (LSUnit12);
		\draw[myArrow] (LSUnit11) -- (z21);
		\draw[myArrow] (LSUnit12) -- (z22);
		
	\end{tikzpicture}
		\caption{Architecture overview of MS-ANFIC. Notation analogous to Fig. \ref{fig:m_anfic_architecture}. Shaded area marks the novel components.}
		\label{fig:ms_anfic_architecture}
	\end{figure*}
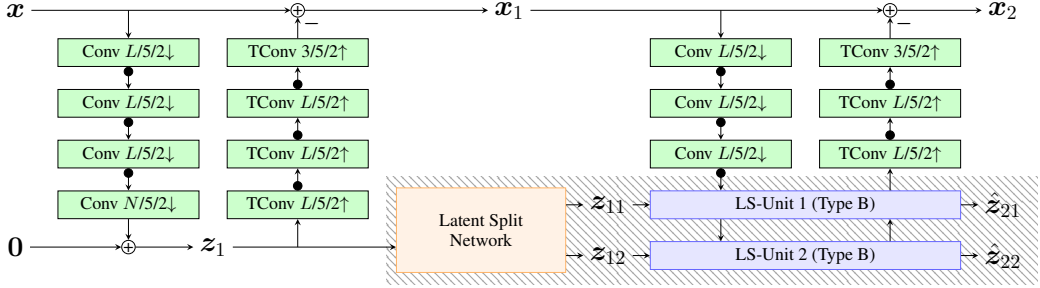
	The first multiscale model, \mbox{M-ANFIC}, applies the multiscale latent space to all layers of the 2-step ANF.
	As shown in Fig. \ref{fig:m_anfic_architecture}, two LSUnits per ANF layer are connected to each of the two autoencoding transforms, which allows to transmit the latent space on two different scales.
	The LSUnits of the first autoencoding transform are of \mbox{type A}, whereas the LSUnits of the last ANF layer are of \mbox{type B}.
	Comparing the structure of the two types in Figs. \ref{fig:lsunit_A} and \ref{fig:lsunit_B} shows, that \mbox{type A} contains solely a downsampling of factor two in the encoding and a upsampling of the same factor in the decoding step.
	The latent space is computed by a convolutional layer without downsampling followed by the masking according to the externally applied mask.
	Besides all previously mentioned components, the LSUnit (Type B) further contains the quantization and the conditional hyperprior network with autoregressive context model to calculate the parameters of the GMM.
	
	\subsection{MS-ANFIC}
	\label{subsec:half_rdoanfic}
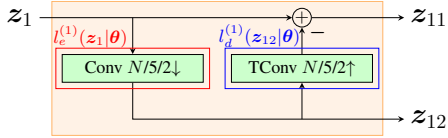
\begin{figure}
		\centering
			\begin{tikzpicture}[
		dataBlob/.style={rectangle, anchor = center},
		Layer/.style={rectangle, draw, anchor = center, fill=green!20, minimum width=2.5cm},
		Attention/.style={rectangle, draw, anchor = center, fill=yellow!20},
		Activation/.style={rectangle, draw, anchor = center, fill=blue!20},
		Misc/.style={rectangle, draw, anchor = center, fill=black!20},
		Code/.style={, draw, anchor = center, fill=black!20},
		bg/.style={rectangle, anchor = center, anchor = north west,},
		myArrow/.style={-stealth},
		myArrowG/.style={*-stealth},
		myArrowL/.style={open diamond-stealth},
		myArrowRE/.style={diamond-stealth},
		myArrowR/.style={stealth-},
		label/.style={rectangle, anchor = center},
		operator/.style={circle ,draw, anchor = center, minimum size = 0.2cm, inner sep = 0pt},
		Network/.style={rectangle, anchor = north west, fill = blue!10, draw = blue!60},
		]
		
		\node (inputZ) [dataBlob] at (0, 0)  {\Large$\bm{z}_{1}$};
		\node (dummy1) [label, right of=inputZ, node distance=2cm] {};
		\node (down) [Layer, below of=dummy1, node distance=0.9cm] {Conv $N$/5/2$\downarrow$};
		\node (plus) [operator, right of=dummy1, node distance=3cm] {$+$};
		\node (minus1) [label, below right=-0.1cm and -0.1cm of plus] {$\boldsymbol{-}$};
		\node (up) [Layer, below of=plus, node distance=0.9cm] {TConv $N$/5/2$\uparrow$};
		\node (dummy2) [label, below of=up, node distance=0.9cm] {};
		\node (outputZ1) [dataBlob, right of=plus, node distance=2.25cm] {\Large$\bm{z}_{11}$};
		\node (outputZ2) [dataBlob, below of=outputZ1, node distance=1.8cm] {\Large$\bm{z}_{12}$};
		
		\begin{pgfonlayer}{background}
			\node (bg) [bg, draw=orange!60, fill=orange!10, fit=(up) (down) (dummy1) (dummy2), inner sep=5pt] {};
		\end{pgfonlayer}
	
		\draw[myArrow] (inputZ) -- (plus);
		\draw[myArrow] (plus) -- (outputZ1);
		\draw[myArrow] (dummy1.center) -- (down);
		\draw[myArrow] (down) |- (outputZ2);
		\draw[myArrow] (dummy2.center) -- (up);
		\draw[myArrow] (up) -- (plus);
		
		\node (le1) [draw=red, fit=(down)] {};
		\node (le1l) [label, above, text=red, xshift=0.62cm, yshift=-0.1cm] at (le1.north west) {\small$l_e^{(1)}(\bm{z}_1|\bm{\theta})$};
		\node (ld1) [draw=blue, fit=(up)] {};
		\node (ld1l) [label, above, text=blue, xshift=0.62cm, yshift=-0.1cm] at (ld1.north west) {\small$l_d^{(1)}(\bm{z}_{12}|\bm{\theta})$};
		
	\end{tikzpicture}
		\caption{Latent split network. Notation analogous to Fig. \ref{fig:m_anfic_architecture}.}
		\label{fig:lssplit}
	\end{figure}

	As second model, we propose \mbox{MS-ANFIC} a multiscale ANFIC with latent split network.
	In contrast to the \mbox{M-ANFIC} architecture, the multiscale latent space is only adapted for the final ANF layer, where the latent representation is transmitted.
	This choice is motivated by the intention to use the different scales mainly in the course of transmission and not for the feature generation.
	As shown in Fig. \ref{fig:ms_anfic_architecture}, the first ANF layer is identical to the original ANFIC architecture, whereas the final ANF layer uses the LSUnits (Type B), like \mbox{M-ANFIC}, to adopt the multiscale latent.
	To transform the single-scale latent derived from the first ANF layer into a multiscale representation an ANF-based latent split network, as shown in Fig. \ref{fig:lssplit}, is used.
	
	The two-scale latent can be computed as:
	\begin{equation}
		\begin{aligned}
			\bm{z}_{12} &= l_e^{(1)}(\bm{z}_1|\bm{\theta})\\
			\bm{z}_{11} &= \bm{z}_1 - l_d^{(1)}(\bm{z}_{12}|\bm{\theta}) \text{,}
		\end{aligned}
		\label{eq:lssplit}
	\end{equation}
	where $\bm{z}_1$ is the single-scale latent generated by the first ANF layer, and $\bm{z}_{11/12}$ are the computed multiscale representations.
	The functions $l_e^{(1)}$ and $l_d^{(1)}$ are learnable functions with the parameter set $\bm{\theta}$.
	The design of the latent split network could also be extended to more than two hierarchy levels, without limiting the invertibility of the transformation.
	
	\section{Experiments and Results}
	\label{sec:results}
	\subsection{Parametrization}
	\label{subsec:param}
	The original ANFIC architecture used with $L=128$ a smaller number of channels in the transform layers, compared to the $N=320$ channels in the latent space.
	Directly adapting these parameters for our multiscale models might favor the use of the first LSUnit over deeper units.
	Therefore, we used $L=N=192$ channels, both for the transform layers and the latent space.
	We performed an ablation study, to ensure that differences in performance do not come from this change in the channel dimensions.
	Thus, we also retrained the original ANFIC with this $L=N=192$ parametrization.
	As shown in Sec. \ref{subsec:rdperformance}, ANFIC achieves comparable results for both configurations.
	
	\subsection{Training}
	\label{subsec:training}
	\begin{table}
		\caption{Training schedule giving the epoch until which a set of learning rate (lr), $\lambda_2$ and mask is used.\\}
		\centering
		\begin{tabular}{c|c|c|c}
			Epoch & lr & $\lambda_2$ & Mask \\
			\midrule
			30 & 1e-4 & 0.1 & Rand \\
			100 & 1e-4 & 0.1 & Var \\
			130 & 1e-4 & $\left\{\begin{array}{c}
				0.1\\
				0.05\\
				0.02\\
				0.01\\
				0.005\\
				0.002
			\end{array}\right\}$ & Var \\
		\end{tabular}
		\label{tab:train_schedule}
	\end{table}
	Similar to the original ANFIC, we used the vimeo-90k \cite{Xue2019} data set to train our models, by taking a random frame of each sequence per epoch cropped to a size of 256\texttimes256.
	The Adam optimizer with standard parameters \cite{Kingma2014} has been used with the training schedule summarized in Tab. \ref{tab:train_schedule}.
	Similar to RDONet \cite{Brand2022b}, we start with random masks first and switch to variance-based masks after 30 epochs.
	We used MSE as our distortion metric $d(\cdot, \cdot)$ and trained the network with the following loss function:
	\begin{equation}
		L = r + \lambda_1 \left|\left|\bm{x}_2\right|\right|_2^2 + \lambda_2 d(\hat{\bm{x}}, \bm{x}) \text{.}
		\label{eq:loss}
	\end{equation}
	Besides the rate $r$ and the MSE distortion metric, an additional loss term forces the residual $\bm{x}_2$ to approximate zero.
	Identical to the original ANFIC, the weighting factor for this term is calculated as $\lambda_1 = 0.01 \cdot \lambda_2$.
	
	\subsection{Rate Distortion Performance}
	\label{subsec:rdperformance}
	We performed our test on the TECNICK \cite{Asuni2014} data set, which contains 100 high quality images of size 1200\texttimes1200.
	The averaged rates are reported in terms of bits per pixel (bpp) and the quality in PSNR-RGB and MS-SSIM \cite{Wang2003}.
	Additionally, we compute the Bj{\o}ntegaard delta \cite{Bjontegaard2001} (BD) rates for both quality metrics.
	Besides our two models \mbox{M-ANFIC} and \mbox{MS-ANFIC}, we give the results of the ANFIC models with $L=128$ and $N=320$, the retrained ANFIC-192 with the parametrization $L=N=192$ and the intra coding mode of the VVC reference implementation VTM in version 18.2.
	For the latter, the PNG images have been converted into 10-bit YUV files with 444 color format. The decoded output is transformed back into PNG files and compared using PSNR-RGB and MS-SSIM metrics.
	The BD-rates, in Tab. \ref{tab:bd_results}, indicate that our \mbox{MS-ANFIC} model performs best, both in terms of PSNR-RGB and MS-SSIM, followed by our \mbox{M-ANFIC} model.
	\begin{table}
		\caption{Bj{\o}ntegaard delta rates on the TECNICK data set for PSNR-RGB and MS-SSIM with ANFIC as anchor. Best model is highlighted bold.\\}
		\centering
		\begin{tabular}{c|cc}
							& PSNR-RGB 			& MS-SSIM\\
			\midrule
			ANFIC (anchor)	& $0\%$ 			& $0\%$\\
			VTM-444			& $0.36\%$ 			& $-0.69\%$\\
			ANFIC-192		& $-2.47\%$ 		& $-6.63\%$\\
			M-ANFIC	(ours)	& $-6.22\%$ 		& $-7.93\%$\\
			MS-ANFIC (ours)	& $\bm{-7.47\%}$ 	& $\bm{-9.06\%}$\\
		\end{tabular}
		\label{tab:bd_results}
	\end{table}
	\begin{figure}
		\centering
		\includegraphics[width=0.85\columnwidth]{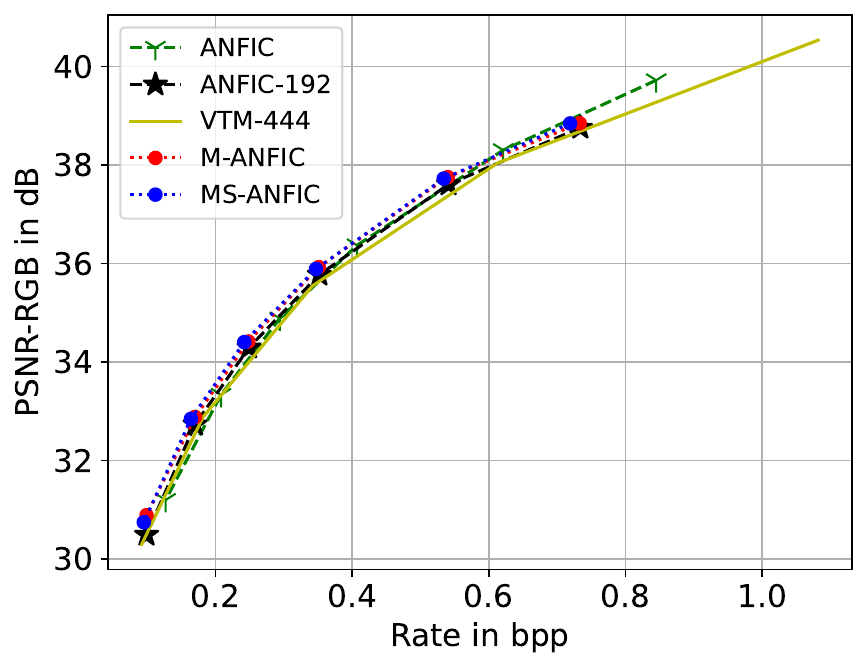}
		\caption{Objective quality evaluation on the TECNICK data set using the PSNR-RGB metric.}
		\label{fig:results_tecnick_psnr}
	\end{figure}
	The results for PSNR-RGB in Fig. \ref{fig:results_tecnick_psnr} show that the ANFIC and ANFIC-192 are on par with the traditional VVC codec.
	Compared to the ANFIC model, our multiscale models \mbox{M-ANFIC} and \mbox{MS-ANFIC} can improve the compression performance by savings of 6.22\% and 7.47\% BD-rate, respectively.
	The \mbox{MS-ANFIC} model with latent split network achieves even better results than \mbox{M-ANFIC}, although the number of parameters are reduced by more than 7\% from \mbox{$\sim$43.0 M} to \mbox{$\sim$39.7 M}.
	Moreover, the results for ANFIC-192 prove, that the gains of our multiscale models is only partially related to the reparametrization of $L=N=192$.
	Compared to ANFIC, ANFIC-192 achieves better compression results for lower rates and negligibly worse for high quality, which is caused by the lower number of channels in the latent representation and results in averaged bit rate savings of 2.47\%.
	Similar to \mbox{MS-ANFIC}, this is not related to the number of parameters as they have been slightly reduced from \mbox{$\sim$22.7 M} for ANFIC to \mbox{$\sim$21.4 M}.
	
\begin{figure}
		\centering
		\includegraphics[width=0.85\columnwidth]{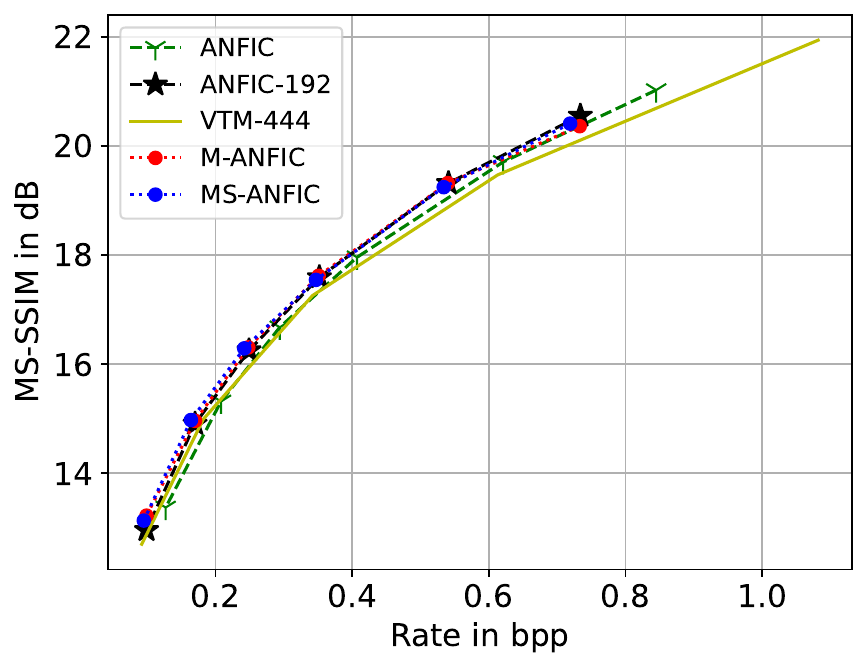}
		\caption{Objective quality evaluation on the TECNICK data set using the MS-SSIM metric. Note that all models were still trained exclusively on MSE.}
		\label{fig:results_tecnick_msssim}
	\end{figure}

	Even though the models are optimized for MSE, we also give the results for MS-SSIM in Fig. \ref{fig:results_tecnick_msssim}.
	We report the MS-SSIM in dB using the formula:
	\begin{equation}
		\text{MS-SSIM}_{\text{dB}} = -10 \log_{10}(1 - \text{MS-SSIM}) \text{.}
		\label{eq:msssim_db}
	\end{equation}
	On a par with the results for the PSNR-RGB metric, the original ANFIC achieves comparable performance to the VTM implementation, whereas they are surpassed both by the reparametrized ANFIC model and our \mbox{M-ANFIC} and \mbox{MS-ANFIC} models.
	Here, we are able to achieve BD-rate savings up to 9.06\% for \mbox{MS-ANFIC} over the original ANFIC implementation.
	
	As our masks for the hierarchical latent space are derived using a variance-based decision, they inherently contain information on the amount of structure present in the different image regions.
	Evidently, the network can exploit this information.
	Based on the knowledge in which level of the latent space an image region was transferred, the hierarchical models can adapt the amount of details generated in the reconstructed image, which leads to a noteworthy improvement in terms of MS-SSIM.
	
	\section{Conclusion}
	\label{sec:conclusion}
	In this paper, we proposed two novel architectures, which extend the concept of an ANF-based image compression network with a hierarchical latent space.
	This multi-scale latent space adds an additional degree of freedom during inference, as the hierarchy level for the different image areas can be set by a mask.
	Thus, the bit rate can be adaptively allocated to more important image areas, e.g. in scenarios like image coding for machines \cite{Fischer2023}, where high quality of certain regions of interest is advantageous.
	Our two models are modified versions of the ANFIC architecture.
	We redesigned the latent space of ANFIC by adding hierarchical LSUnits, adapted based on RDONet, and developed an invertible latent split network for our \mbox{MS-ANFIC} model, which can derive a multiscale representation from a single-scale latent.
	
	The adoption of a multiscale latent space can noticeably improve the compression performance compared to the single-scale ANFIC.
	With the usage of a multiscale latent space for the final ANF layer, see the \mbox{MS-ANFIC} model, we are able to save on average more than 7\% bit rate at same PSNR and 9\% at same MS-SSIM.
	Our results prove, that the additional flexibility due to the multiscale principle can enhance the performance not only for compressive autoencoders but also in ANF-based architectures. 
	These outcomes show, that improvement of learning-based image compression is not limited to developing better transformations and entropy models.
	Also the overall architecture has a major impact in the compression performance. 
	
	\bibliographystyle{IEEEbib}
	\bibliography{windsheimer.bib}
	
\end{document}